\documentclass[10p,conference]{IEEEtran}
\usepackage{graphics}                 
\usepackage{graphicx}
\usepackage[cmex10]{amsmath}
\usepackage{amssymb, amsfonts}
\usepackage{algorithmic}
\usepackage{algorithm}
\usepackage{cite}
\usepackage{verbatim}
\usepackage{url}
\ifCLASSOPTIONcompsoc
\usepackage[tight,normalsize,sf,SF]{subfigure}
\else
\usepackage[tight,footnotesize]{subfigure}
\fi
\usepackage{subfigure}
\usepackage{slashbox}

\begin{document}
%
\title{Cooperation Enforcement for Packet Forwarding Optimization in Multi-hop Ad-hoc Networks}

\author{\IEEEauthorblockN{Mohamed-Haykel Zayani\IEEEauthorrefmark{1}, Djamal Zeghlache\IEEEauthorrefmark{1}}
\IEEEauthorblockA{\IEEEauthorrefmark{1}Lab. CNRS SAMOVAR UMR 5157,
Telecom SudParis, Evry, France} \IEEEauthorblockA{Emails:
\{mohamed-haykel.zayani, djamal.zeghlache\}@telecom-sudparis.eu}
 }


%





\maketitle

\begin{abstract}
Ad-hoc networks are independent of any infrastructure. The nodes are
autonomous and make their own decisions. They also have limited
energy resources. Thus, a node tends to behave selfishly when it is
asked to forward the packets of other nodes. Indeed, it would rather
choose to reject a forwarding request in order to save its energy.
To overcome this problem, the nodes need to be motivated to
cooperate. To this end, we propose a self-learning repeated game
framework to enforce cooperation between the nodes of a network.
This framework is inspired by the concept of ``The Weakest Link" TV
game. Each node has a utility function whose value depends on its
cooperation in forwarding packets on a route as well as the
cooperation of all the nodes that form this same route. The more
these nodes cooperate the higher is their utility value. This would
establish a cooperative spirit within the nodes of the networks. All
the nodes will then more or less equally participate to the
forwarding tasks which would then eventually guarantee a more
efficient packets forwarding from sources to respective
destinations. Simulations are run and the results show that the
proposed framework efficiently enforces nodes to cooperate and
outperforms two other self-learning repeated game frameworks which
we are interested in.
\end{abstract}

\IEEEpeerreviewmaketitle


\section{Introduction}
In wireless Ad-hoc networks, nodes are self-organizing and
autonomous. They manage their own resources and make their own
decisions. In order to maintain network connectivity, each node has
to forward packets of other nodes. However, since they are known to
have limited battery resources, these nodes usually tend to be
non-cooperative. Indeed, they might sometimes reject forwarding
requests in order to save their proper energy. Thus, nodes are
reluctant to participate in routing which may lead the network
connectivity to break down. It is then necessary to provide a
mechanism that enforces cooperation between nodes and maintains the
network connectivity. The problem of forwarding packets in a
non-cooperative Ad-hoc network is widely studied, as we highlight
afterwards, and many approaches have been proposed. The nodes act
selfishly and tend to maximize their own benefits, thus, most of
these studies rely on game theory \cite{Fudenberg1991} which is a
suitable tool to deal with complex interactions between network
nodes. From this perspective, the approaches can be classified into
two categories depending on the mechanism used to enforce
cooperation level between nodes. In the first category, the
propositions use the virtual payment scheme. Zhong et al.
\cite{Zhong2003} have proposed Sprite, a credit-based system that
makes incentives for nodes to cooperate. Eidenbenz et al. have
designed COMMIT \cite{Eidenbenz2008}, a routing protocol based on
payment with virtual currencies. The requested intermediate nodes
will perceive a compensation that is related to their residual
energy level. Ad Hoc-VCG \cite{Anderegg2003} and incentives modeling
advanced by Crowcroft et al. \cite{Crowcroft2004} also belong to the
first category. In the second one, the approaches employ mechanisms
to enforce and maintain cooperation between nodes communities. Some
works use reputation-based mechanisms. In instance, Kwon et al.
\cite{Kwon2010} who have formulated a Stackelberg game where two
nodes sequentially estimate the willingness of each other and decide
to cooperate or not according to the opponent reputation score.
Also, Buchegger and Le Boudec \cite{Buchegger2002a,Buchegger2002b}
have defined mechanisms taking into consideration reputation system.
Other solutions aim to maintain cooperation by considering
punishment threat. In this case, Marti et al. \cite{Marti2000} have
defined ``watchdog" and ``pathrater" techniques that improve
throughput by excluding misbehaving nodes. Felegyhazi et al.
\cite{Felegyhazi2003} have proposed a scheme that enables the nodes
to reach the Nash Equilibrium, under topological conditions (i.e.
dependence graph), relying on the ``Tit-For-Tat" punishment. Altman
et al. \cite{Altman2005} have highlighted the ``aggressive"
punishment in \cite{Felegyhazi2003} and have proposed milder
punishment mechanism which guaranties a Nash Equilibrium and helps
nodes to consume less energy. Han et al. \cite{Han2005} have
advanced a self-learning repeated framework based on punishment that
determines the nodes optimal packet forwarding probabilities to
maintain network connectivity. Pandana et al. \cite{Pandana2008}
have considered the same aim as in \cite{Han2005} and have designed
three learning schemes with a punishment mechanism under
perfect/imperfect local observation and dependence graph conditions.

In this paper, we propose a self-learning repeated game framework
inspired by ``The Weakest Link" TV game. In our approach, the set of
nodes forming a route are considered as the candidates of the chains
in the TV game. Each node forwarding probability can be seen as good
answer probability for each candidate. Indeed, the maximization of
global collective gains depends strongly on the cooperation between
the candidates involved in the game. Thereby, we adopt the TV game
concept to design our scheme with the objective of motivating nodes
to create the longest chains and maximizing their utility values.
This would increase the probability that the packets are delivered
to the destination and then would optimize packets forwarding.
Moreover, the framework relevance lies in the repeated game that
enforces collaboration between nodes and a learning scheme that tend
to reach better cooperation level. We consider also a punishment
mechanism that would discourage nodes from acting selfishly. To
asses the efficiency of our proposal, we compare our proposal to two
other self-learning repeated game schemes proposed in \cite{Han2005}
and \cite{Pandana2008}.

The remainder of the paper is organized as follows: In Section II,
the proposed model based on the ``Weakest Link" TV game principle is
illustrated. The self-learning repeated game framework and
punishment scheme are then presented in Section III. Section IV
details simulations scenarios used to evaluate our proposal and
analyzes the obtained results. Finally, concluding remarks are given
in Section V.


\section{System Model And Problem Formulation}
To optimize packet forwarding in Ad-hoc networks, we strongly
believe that applying the concept of the ``Weakest Link" TV game is
an interesting solution. We explain, firstly, the concept of the
game that we want to reproduce and then we detail the proposed
model.

\subsection{The ``Weakest Link" TV Game Principle}
Weakest Link TV game is a game where a group of candidates try to
answer correctly, relying on their knowledge, to the questions asked
by the TV host. They aim to gather the highest amount of money
through successive rounds. In each round, the players try to form a
chain of nine correct answers to reach the highest gain. Before
answering to a question, a candidate has the possibility to save the
collected gain, provided by good answers of precedent candidates, by
saying ``Bank". It is obvious that the longer the chain is, the
higher the gain gets. Nevertheless, a player who gives a wrong
answer to a question and did not save the collected gains shall
break the good answers chain and reset the gain to zero. Secondly,
if the candidates save rapidly the collected gains, the chains will
be too short to reach important amounts of money. Thus, answering
wrongly or saving collected money return the chain counter to zero.
The candidates have to avoid being frequently in these situations if
they want to maximize their gains. The earnings scale expresses the
potential round gain according to correct answers chain length. For
example, if there are four good answers and the current player
decides to save collected money, the total gains grow with the
amount that corresponds to the chain length. Then, the candidates
try to create another chain of good answers. Therefore, the key
parameter that influences the maximization of earnings is the
probability of giving a correct answer.

\subsection{The Proposed Model: Analogy with the ``Weakest Link" TV
Game} Our objective is to define an approach that enforces
cooperation in a distributed way. The model we propose is inspired
by the principle of the TV game. We note interesting analogies
between ``The Weakest Link" TV game and the Ad-hoc network. The
nodes are assumed to be the candidates of the game and forwarding a
packet from a node to the next hop is considered as a good answer.
We believe strongly that the TV game concept can be used to
encourage nodes to cooperate and therefore to optimize packet
forwarding. Hence, the nodes, along a route, aim to create the
longest chain of successful forwarded packets to get better utility.
Despite of the TV game, when a chain of forwarded packet is broken,
only the nodes that form the chain and the node which saves the
gains will be rewarded. The set of all nodes composing the route is
rewarded by the collected gains only if the packet reaches
destination. To formulate the expected utility of a node in a route,
we propose expressions inspired by \cite{Ben-Ameur2006}. Let
$\alpha_{i}$ and $\beta_{i}$ be the forwarding and the saving gain
(with chain breaking) probabilities, respectively, for each node
$i$. We assume in our work that $\alpha_{i}=1-\beta_{i}$. Given a
route $R$ with $N$-1 hops ($N$ nodes), we define in Eq. (1) the
average gain that a node $i$ can expect when it plays the role of
the $n^{th}$ link of the chain. Let $S(i)$ be the next hop of node
$i$ in route $R$. We mean by $C[n]$ the collected earnings when a
chain of ($n$-1) successful transmissions is transformed in currency
and by $F$ the cost of forwarding other's packets. Considering the
vector of forwarding probabilities $\alpha$, the utility of a node
$i$ is given by:
\begin{equation}
U_{i}^{R}(n,\alpha)=\left\{\begin{matrix} U_{S(i)}^{R}(1,\alpha) &
if \:
n=0 \\ \\
(1-\alpha_{i}).C[n]+\\
\alpha_{i}.(U_{S(i)}^{R}(n+1,\alpha)-F) & if \: 0<n<N\\
\\ C[n] & if \: n=N
\end{matrix}\right.
\label{form1}
\end{equation}

When the node $i$ is an intermediate node and given that it wants to
maximize its gains, it has to choose between saving the collected
currency or increasing the chain length (relying on the cooperation
of the successor(s) to maximize benefits despite of the cost of
forwarding). We assume that the source of the packet (i.e. $n$=0)
will send it with the probability equal to 1. Subsequently, the
gains depend on the decisions of the next hops. In addition, when
the packet reaches the destination (i.e. $n$=$N$), the node will
save the collected gains with a probability equal to 1. In the
latter case, the chain has the length of the route and the gains are
maximized.

Therefore, we can formulate this problem as a non-cooperative game
where each node will adjust its forwarding probability in order to
maximize its own utility. A node $i$ can belong to more than one
route, its own utility is then the sum of each route utility, called
$U_{i}$. To solve this problem, it is necessary to find the Nash
Equilibrium of the game.

\textbf{\emph{Definition1:}} The Nash Equilibrium is some strategy
set $\alpha^{*}$ for all nodes, such that for each node $i$, the
following condition is verified:
\begin{equation}
U_{i} ( \alpha_{i}^{*},\alpha_{-i}^{*} ) \geq U_{i} (
\alpha_{i},\alpha_{-i}^{*}), \forall i, \forall \alpha_{i}\in \left
[0,1 \right ] \label{form2}
\end{equation}
Where
$\alpha_{-i}^{*}=(\alpha_{1}^{*},\ldots,\alpha_{i-1}^{*},\alpha_{i+1}^{*},\ldots,\alpha_{N}^{*}).$

This condition emphasizes that no node can increase its utility by
operating a unilateral change of its forwarding probability, while
all other nodes play the Nash Equilibrium strategy. However, as
mentioned in several works as \cite{Altman2005,Han2005,Pandana2008},
this equilibrium matches with the strategy where $\alpha_{i}=0,
\forall i$. To avoid a poor network performance, we propose a
self-learning repeated game framework, inspired by the concept of
``The Weakest Link" TV game that we call The Weakest Link scheme.
The objective of this framework is to enforce cooperation between
nodes through learning and punishing threat mechanisms.

\section{Self-Learning Repeated Game Framework and Punishment Mechanism}
\begin{figure}[!tb]
\begin{center}
  \includegraphics[width=0.4\textwidth]{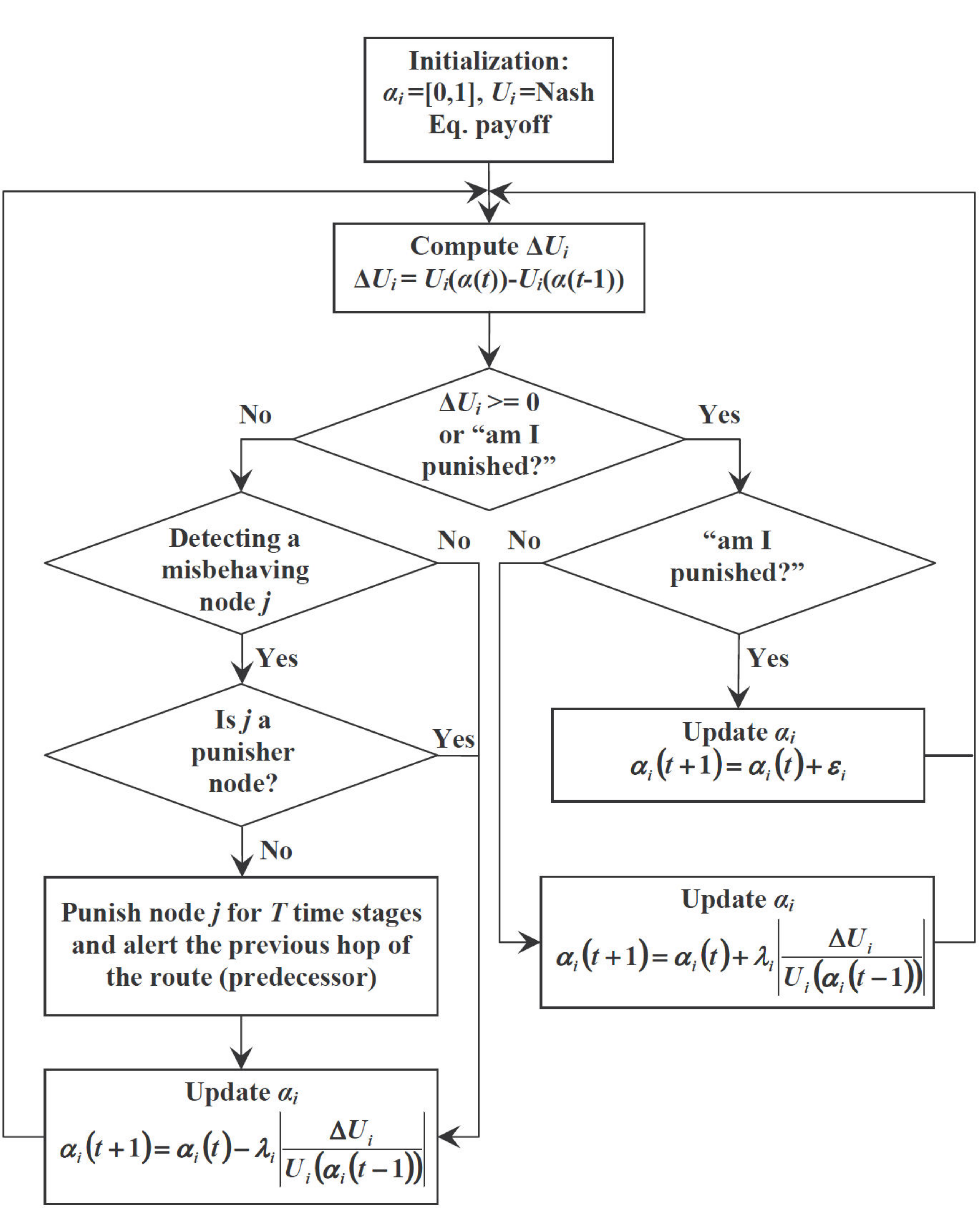}
  \caption{The self-learning repeated game flowchart}
  \label{Flowchart}
\end{center}
\end{figure}
As the Nash Equilibrium corresponds to a non-cooperative strategy,
it is more suitable to design cooperation under repetitive game.
Applying repeated game scheme match perfectly with our proposed
model. Indeed, in each game step, along each route, the nodes try to
maximize their utilities and would choose to cooperate. In this
paper, we consider an infinite repeated game, where the game
duration is unknown to all nodes. Relying on the Folk theorem
\cite{Fudenberg1986}, the outcome of an infinitely repeated game can
give better payoffs than those that can be obtained with Nash
Equilibrium, especially when permanent punishment threat obliges
selfish nodes to be more cooperative. Therefore, we propose a
repeated game that enforces cooperation and maintains it through a
designed punishment mechanism to encounter misbehaving forwarders.
The framework we propose is presented by the flowchart in the Fig.
\ref{Flowchart}.

In the initialization step, all nodes are more or less selfish. They
can set their forwarding probabilities to 0 (the Nash Equilibrium
strategy played with one stage game). We assume that the routes are
determined with a routing protocol and that each node knows all the
routes to which it belongs, as considered in \cite{Han2005,
Pandana2008}. Then, the nodes start playing repeated game strategy.
Note that they are rational and want to make benefits. Thus, at each
step, each node learns through its utility the cooperation level of
other nodes and adjusts its forwarding probability following to the
others learnt behavior. It is possible that a node deviates from
cooperation, then a punishment scheme is designed to discourage
misbehaving nodes and to ask them to be more cooperative in the
future. It is applied as soon as a selfish behavior is detected and
subsequently the framework satisfies the Folk theorem.

\subsection{The punishment mechanism}
In this section, we present the punishment mechanism through a
simple scenario. We assume that $A$ and $B$ are two successive nodes
along a route. We suppose that the node $B$ is a misbehaving node.
The node $A$ as one among the ``closest" nodes to node $B$ (i.e. the
predecessor of $B$) is designed to punish it (if $B$ rejects the
request of $A$). We assume that the node $A$ is able to detect the
lack of cooperation of $B$ (able to distinguish between a packet
drop and a packet loss); the node $A$ can conclude, by listening the
channel, that node $B$ is misbehaving when it does not forward the
packet to the next node. To punish the selfish node $B$, the node
$A$ fixes its forwarding probability to 0 when the packet has to
pass through node $B$. In this case, the node $B$ will not be able
to receive any packet from node $A$. Thereafter, the node $B$ will
be excluded from all chains in which its predecessor is in
punishment mode. This punishment cancels the node $B$ benefits for a
period $T$ and enforces it to cooperate (as its utility decreases).
To avoid the propagation of the punishment mode over all nodes, when
the node $A$ is designated to punish the node $B$, the former one
informs its predecessor about the execution of the punishment. Then,
the punishment act is not interpreted as a deviation. It is
important to mention that we assume that the nodes are not
malicious.

\subsection{The self-learning repeated game framework description}
At each step of the repeated game, each node compares its current
utility value with the former value. If the current utility is
better, a cooperation enforcement is concluded. Thereby, the
forwarding probability is increased proportionally with the
enhancement of the utility to promote the cooperation level. The
upgrade of the cooperation level is also led by the coefficient
$\lambda_{i}$. It expresses the node sensitivity to the cooperation
enforcement. However, when the current utility drops, it is analyzed
as a come back to selfishness. Thus, the forwarding probability will
decrease (proportionally with the difference). Analogically with
``The Weakest Link" TV game, a candidate chooses to break the chain
and insures gains if he notices that the following candidate tends
to make wrong answers. Hence, cooperative nodes are sensitive to the
behavior of the other nodes. As described in the punishment scheme,
a node checks if its successor deviates. This deviation can be the
result of punishment and an announcement is made to avoid
selfishness propagation. The deviation without any notification is
considered as a selfish behavior and then the punishment procedure
is applied on the misbehaving node. Indeed, during a period $T$, no
packet from its punishing predecessor reaches it. This causes a
dramatic utility decrease. Therefore, a punished node is encouraged
to be more cooperative in order to avoid longer penalization.
Indeed, it increases its forwarding probabilities by a step equal to
$\varepsilon_{i}$. To make possible the cooperation enforcement,
another assumption must be considered. In fact, if the maximum gains
that can be collected are lower than the forwarding cost, the nodes
would not forward any packet even under punishment threat. Finally,
it is important to mention that the proposed scheme work well when
the mobility is moderate. In other words, transferring all packets
on a route must be faster than route breakage due to mobility. We
take into consideration this assumption in our simulations.

This framework aims to find the longest chains on a route between
source and destination nodes, and routes are provided by a routing
protocol to all nodes. If a route is not available at any node in
the route, the routing tables must be updated. Moreover, the
forwarding probability can be used by the routing protocol to
determine better routes or to update routes in order to avoid
misbehaving nodes to be a part of a route. Also, it is important to
mention that this framework is adapted to scenarios where some nodes
can either be only sources and/or destinations of packets as a
source node forwards its packets to the next hop with a probability
equal to 1 (only if the source node apply the punishment mechanism)
and a destination node is not involved in the forwarding process.

\begin{figure}[tb]
\begin{center}
  \includegraphics[width=0.3\textwidth,angle=270]{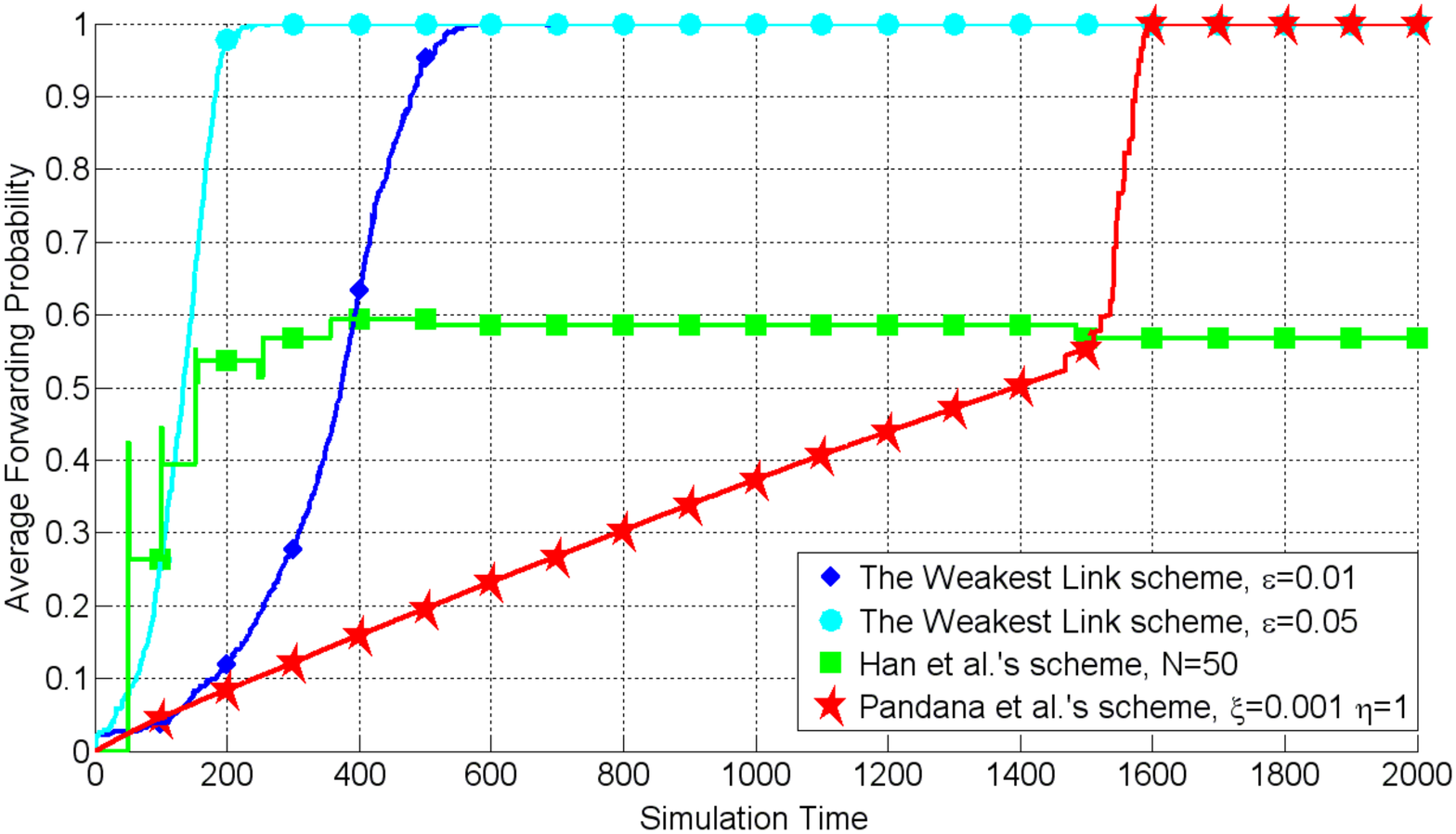}
  \caption{The evolution of the average forwarding probability for different self-learning repeated game schemes in the case of the ring network}
  \label{Comparison1}
\end{center}
\end{figure}
\begin{figure}[tb]
\begin{center}
  \includegraphics[width=0.3\textwidth,angle=270]{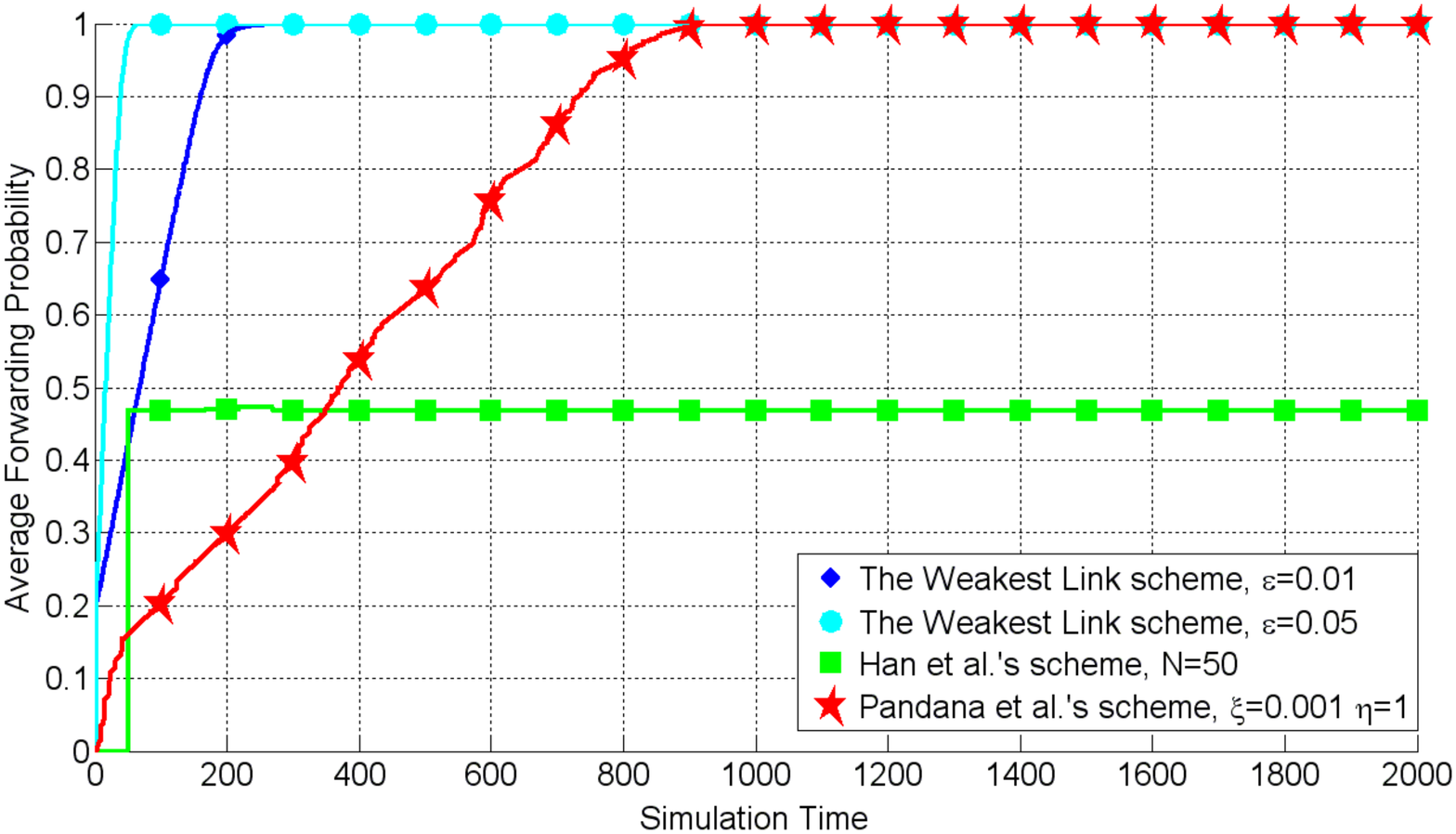}
  \caption{The evolution of the average forwarding probability for different self-learning repeated game schemes in the case of a random network}
  \label{Comparison2}
\end{center}
\end{figure}

\section{Performance Evaluation and Simulations Results}
In this section, we evaluate the performances of the proposed
approach through two scenarios: the widely used ring network and the
random network. We compare our proposal to two other approaches
which have inspired us: the scheme proposed by Han et al.
\cite{Han2005} and the learning through flooding algorithm designed
by Pandana et al. \cite{Pandana2008}. We implement our framework and
do scenarios simulations on MATLAB 7.8.0.

Firstly, we consider a ring network of 25 nodes. The distance
between any source and destination is $N$ hops (if a node $i$ is the
source, node $mod(i+N,25)$ is the destination). For each
intermediate node, it is imperative that the forwarding cost must be
less important than the maximum gain along a route. By the way, the
nodes must have benefits in order to maintain a cooperative
behavior. In our simulations, each successful forwarding increments
by 1 the gain corresponding to the formed chain. In this scenario,
we fix $N$ to 6. We consider that one node can be a source or a
destination at most one time and an intermediate node at most five
times. The nodes initialize their forwarding probabilities at 0
(i.e. the Nash Equilibrium strategy). We choose also to fix the
forwarding cost to 3 (i.e. to make benefits possible) and the period
of punishment $T$ to 3 time steps.

We represent in Fig. \ref{Comparison1} the evolution of the average
forwarding probabilities for the considered self-learning repeated
game frameworks over 2000 time steps. For our proposal, we depict
two different scenarios and each one is characterized by a specific
value of $\varepsilon$ (where $\varepsilon_{i}=\varepsilon$ for all
nodes and $\varepsilon$ takes respectively the values 0.01 and
0.05). The coefficient $\lambda_{i}$ is fixed to 0.01 for all nodes.
We have also fixed the characteristic parameters of the two other
schemes as mentioned in the plot. We invite the reader to refer to
\cite{Han2005} and \cite{Pandana2008} in order to understand the
meaning of these parameters (we also assume that each two successive
nodes in a route are separated by the same distance in order to
simplify the computation of the utility functions when the scheme of
Pandana et al. is used).

First of all, we remark that the average forwarding probability
observed for the scheme of Han et al. converges to a value that
turns around 0.6 and we note that the updates of this average
probability become less frequent as time goes on. This finding
logically indicates that the cooperation between nodes is limited
even if the average cooperation level is rather substantial. The
punishment mechanism adopted by the framework of Han et al. is
clearly the major cause behind this result and we explain later the
reasons behind that. The average forwarding probability obtained for
the learning through flooding algorithm designed by Pandana et al.
converges to 1 but after too many steps (compared to the Weakest
Link scheme result). In their evaluation, Pandana et al. have chosen
an initial forwarding strategy for nodes (i.e all forwarding
probabilities initialized to 0.5) different from the Nash
Equilibrium strategy (i.e. non-cooperative strategy). This
assumption enables to the nodes of the network that adopt this
scheme to reach more rapidly high cooperation levels but leads to a
skewed evaluation of the algorithm. Regarding the Weakest Link
scheme, the average forwarding probability converges to 1, as the
scheme of Pandana et al. but needs less time to reach high
cooperation levels. We depict the corresponding evolution for two
values of $\varepsilon$: 0.01 and 0.05. We remark that the nodes
become cooperative faster as they are more reactive to the
punishment mechanisms (i.e. higher increase of the forwarding
probability when punishment and utility decrease are detected).
Then, the scenario where $\varepsilon$ is equal to 0.05 highlights a
quicker convergence of the average forwarding probability.

\begin{table}[t]
\renewcommand{\arraystretch}{1}
\caption{Evaluation metrics in the case of the ring network (all
forwarding probabilities initialized to 0)} \label{table1}
\centering \scalebox{0.7}{
\begin{tabular}{|l|c|c|}
\hline
\backslashbox{\bfseries Schemes}{\bfseries Metrics} & \bfseries Avg. PDR @ Dest. & \bfseries Fwd. Pkts / Dlv. Pkts\\
\hline\hline
The Weakest Link ($\varepsilon$=0.01) & 75.19\% & 5.1195 \\
\hline
The Weakest Link ($\varepsilon$=0.05) & 90.30\% & 5.0452 \\
\hline
Pandana et al. & 21.29\% & 5.7362 \\
\hline
Han et al. & 0.23\% & 182.7456 \\
\hline
\end{tabular}}
\end{table}

\begin{table}[t]
\renewcommand{\arraystretch}{1}
\caption{Evaluation metrics in the case of a random network (all
forwarding probabilities initialized to 0)} \label{table2}
\centering \scalebox{0.7}{
\begin{tabular}{|l|c|c|}
\hline
\backslashbox{\bfseries Schemes}{\bfseries Metrics} & \bfseries Avg. PDR @ Dest. & \bfseries Fwd. Pkts / Dlv. Pkts\\
\hline\hline
The Weakest Link ($\varepsilon$=0.01) & 92.94\% & 4.5709 \\
\hline
The Weakest Link ($\varepsilon$=0.05) & 98.21\% & 4.5459 \\
\hline
Pandana et al. & 67.31\% & 4.7061 \\
\hline
Han et al. & 1.62\% & 8.6341 \\
\hline
\end{tabular}}
\end{table}

To support these conclusions, we determine for each scheme the
average packet delivery rate at the destination and the ratio
between the forwarded packets and the delivered packets. Table
\ref{table1} lists the corresponding results for the ring network
scenario. It is important to remind that the routes have a length of
6 hops. Then, in the ideal case when the nodes fully cooperate, each
packet delivered to the destination needs 5 forwards. The found
results reflect the efficiency of our scheme. This efficiency is
tuned by the input parameters. As we show, the choice of the
parameter $\varepsilon$ has an important impact on the convergence
speed of the average forwarding probability to 1. The scheme of
Pandana et al. shows a limited efficiency over the simulation time
because of the low convergence to a satisfying cooperation level.

For the two cited schemes, the nodes become cooperative as time goes
on. The two algorithms rely on efficient punishment mechanisms. They
share the penalty of the misbehaving node instead of all nodes in
the network. The values obtained for the ratios of forwarded packets
over delivered packets (slightly higher to 5) prove the
effectiveness of these solutions to establish cooperation among
nodes (even with some delay for the scheme of Pandana et al.). On
the contrary, when a node that uses the framework of Han et al.
detects a defection, it punishes all other nodes. This reaction
engenders the propagation of the non-cooperative strategies and
dramatically falls down the network performance.

In the same way, we consider the scenario of a random network
consisting of 100 nodes with 1000 source-destination pairs. Fig.
\ref{Comparison2} depicts the evolution of the average forwarding
probability for each scheme and Table \ref{table2} lists the
evaluation metrics for the scenario of the random network (the
average number of forwarders per route in this scenario is 4.535
nodes). The previous observations and interpretations match with the
results obtained for this scenario. We note also that the
performance of the Weakest Link and Pandana et al.'s schemes are
better. This can be explained by higher opportunities to improve the
utility function (i.e. nodes belong to a higher number of routes)
compared to the case of the ring network.

Pandana et al. have defined the utility function of a node as its
transmission efficiency. The transmission efficiency is the ratio of
successful self-transmission power over the total consumed power
(self-transmission and forwarding). We aim to compare our proposal
to the algorithm of Pandana et al. using the cited criterion. We
consider the same scenario of the ring network, as previously, and
we initialize the forwarding probabilities of nodes at 0.5 to be as
close as possible to the simulation inputs considered by Pandana et
al.

\begin{table}[t]
\renewcommand{\arraystretch}{1}
\caption{Evaluation metrics in the case of the ring network (all
forwarding probabilities initialized to 0.5)} \label{table3}
\centering \scalebox{0.7}{
\begin{tabular}{|l|c|c|c|}
\hline
\backslashbox{\bfseries Schemes}{\bfseries Metrics} & \bfseries Avg. PDR @ Dest. & \bfseries Fwd. Pkts / Dlv. Pkts & \bfseries Avg. Trans. Eff.\\
\hline\hline
The Weakest Link ($\varepsilon$=0.01) & 95.24\% & 5.0555 & 0.1617 \\
\hline
The Weakest Link ($\varepsilon$=0.05) & 97.66\% & 5.0278 & 0.1639 \\
\hline
Pandana et al. & 93.84\% & 5.0492 & 0.1581 \\
\hline
\end{tabular}}
\end{table}

\begin{table}[t]
\renewcommand{\arraystretch}{1}
\caption{Evaluation metrics in the case of the random network (all
forwarding probabilities initialized to 0.5)} \label{table4}
\centering \scalebox{0.7}{
\begin{tabular}{|l|c|c|c|}
\hline
\backslashbox{\bfseries Schemes}{\bfseries Metrics} & \bfseries Avg. PDR @ Dest. & \bfseries Fwd. Pkts / Dlv. Pkts & \bfseries Avg. Trans. Eff.\\
\hline\hline
The Weakest Link ($\varepsilon$=0.01) & 99.77\% & 4.4674 & 0.1832 \\
\hline
The Weakest Link ($\varepsilon$=0.05) & 99.31\% & 4.4722 & 0.1829 \\
\hline
Pandana et al. & 93.54\% & 4.525 & 0.1778 \\
\hline
\end{tabular}}
\end{table}

We list in Table \ref{table3} the average packet delivery rate, the
ratio of forwarded packets over delivered packets and the average
transmission efficiency obtained by the Weakest Link scheme and the
learning through flooding algorithm for the ring network scenario
(the same scenario as previously). We note that the two frameworks
highlight high packet delivery rates at destination (i.e. over 93 \%
with better performance for the Weakest Link scheme) and strong
forwarding effectiveness (i.e. ratios of forwarded packets over
delivered packets slightly higher than 5). Regarding the comparison
based on the transmission efficiency, Table \ref{table3} emphasizes
that the Weakest Link scheme (with the different values chosen for
$\varepsilon$) reaches higher average transmission efficiency than
the algorithm of Pandana et al. and then allows nodes to use better
their energy. This finding can be explained be the upper packet
delivery rate at the destination obtained for the Weakest Link
scheme.

We can in addition emphasize that the behavior of nodes under the
Weakest Link scheme enables nodes to have more ``elastic" behavior
towards defections. Thereafter, it is possible to avoid useless
forwards and save energy. In fact, when a node detects a reduction
in its utility function, it decreases its forwarding probability and
then becomes reluctant to cooperate as the delivery of packets is
not accurate. For the algorithm of Pandana et al., the forwarding
probability can either increase or remain the same but never
decreases. Then, the nodes maintain their cooperation level even if
a defection is detected and can uselessly consume their energy.

We compute the same evaluation metrics for the case of a random
network which consists of 100 nodes and 1000 source-destination
pairs. Each route has at average 4.463 forwarders. We list in Table
\ref{table4} the obtained results for this scenario. We note that
the effectiveness of our proposal is verified and that the
transmission efficiency provided by the Weakest Link scheme is
always better than the one of the algorithm of Pandana et al..

We have to mention that Pandana et al. have also proposed another
framework based on utility prediction. It was highlighted in
\cite{Pandana2008} that this latter scheme enables nodes to have
better transmission efficiency. Anyway, the behavior of the average
forwarding probability follows the same one of the learning through
flooding scheme insofar as nodes can only maintain or increase their
forwarding probabilities. Nevertheless, a convenient choice of the
parameter $\varepsilon$ for our scheme enables us to still improve
the network performance if necessary.


\section{Conclusion}
In wireless Ad-hoc networks, nodes are requested to forward traffic.
However, because of limited energy resources, they might refuse to
collaborate in order to save their energy. This can lead to a
significant amount of lost packets and a deterioration of the
network performances.

In order to overcome this problem, we have proposed in this paper a
self-learning repeated game framework that aims to enforce
cooperation between nodes. Our framework is inspired by ``The
Weakest Link" TV game concept. Indeed, the amount of the global
collective gains strongly depends on the cooperation degree between
the candidates involved in the game. The candidates try to form the
longest chain in order to reach the highest gain. Analogically, the
nodes, along a route, would tend to achieve the longest sequence of
successful packet forwarding and therefore assure that the packet
reaches the destination. Our approach is designed as a self-learning
repeated game framework that enables nodes to learn each others
cooperation levels. Therefore, nodes that are in a same route and
that have a high cooperation level may encourage the other nodes of
the route to get more cooperative. For this aim, a punishment
mechanism has been considered. Thereby, misbehaving nodes are
punished and their utility would dramatically decrease. This allows
the network to maintain a relatively satisfying cooperation level.

Simulations have been run and the results have shown that our scheme
is efficient for the ring network scenario as well as for the random
network scenario. It has been also shown that our proposal
outperforms the self-learning repeated game frameworks that we have
been interested in in this work.

As future work, it would be challenging to test our framework on
real testbeds as smart grids. From this perspective, we want to
enhance the Weakest Link scheme framework by considering the
different channel characteristics among nodes, relaxing the
assumption that a node is able to distinguish between a packet drop
and a packet loss and probably taking into consideration the
residual energy for each node as a parameter in the utility
function. All these perspectives would be helpful to design a real
cooperation enforcement framework for multi-hop Ad-hoc networks.



\end{document}